\documentclass[twocolumn,showpacs,prd]{revtex4}
\usepackage{mathrsfs}
\usepackage{longtable,lscape}
\usepackage{txfonts}
\usepackage{amssymb}
\usepackage{indentfirst}
\usepackage{graphicx,,booktabs}
\usepackage{color}
\usepackage{amssymb}

\begin{document}

\title{Charged bottomonium-like states $Z_b(10610)$ and $Z_b(10650)$
and the $\Upsilon(5S)\to \Upsilon(2S)\pi^+\pi^-$ decay}
\author{Dian-Yong Chen$^{1,3}$}\email{chendy@impcas.ac.cn}
\author{Xiang Liu$^{1,2}$\footnote{Corresponding author}}\email{xiangliu@lzu.edu.cn}
\author{Shi-Lin Zhu$^4$\footnote{Corresponding author}}\email{zhusl@pku.edu.cn}
\affiliation{ $^1$Research Center for Hadron and CSR Physics,
Lanzhou University and Institute of Modern Physics of CAS, Lanzhou 730000, China\\
$^2$School of Physical Science and Technology, Lanzhou University, Lanzhou 730000,  China\\
$^3$Nuclear Theory Group, Institute of Modern Physics of CAS, Lanzhou 730000, China\\
$^4$Department of Physics and State Key Laboratory of Nuclear
Physics and Technology, Peking University, Beijing 100871, China}
\date{\today}

\begin{abstract}

Inspired by the newly observed two charged bottomonium-like
states, we consider the possible contribution from the
intermediate $Z_b(10610)$ and $Z_b(10650)$ states to the
$\Upsilon(5S)\to \Upsilon(2S)\pi^+\pi^-$ decay process, which
naturally explains Belle's previous observation of the anomalous
$\Upsilon(2S)\pi^+\pi^-$ production near the peak of
$\Upsilon(5S)$ at $\sqrt s=10.87$ GeV [K.F. Chen {\it et al}.
(Belle Collaboration), Phys. Rev. Lett. {\bf 100}, 112001 (2008)].
The resulting $d\Gamma(\Upsilon(5S)\to
\Upsilon(2S)\pi^+\pi^-)/dm_{\pi^+\pi^-}$ and
$d\Gamma(\Upsilon(5S)\to \Upsilon(2S)\pi^+\pi^-)/d\cos\theta$
distributions agree with Belle's measurement after inclusion of
these $Z_b$ states. This formalism also reproduces the Belle
observation of the double-peak structure and its reflection in the
$\Upsilon(2S)\pi^+$ invariant mass spectrum of the $\Upsilon(5S)\to
\Upsilon(2S)\pi^+\pi^-$ decay.

\end{abstract}

 \maketitle

Very recently, the Belle Collaboration announced the first
observation of two charged bottomonium-like states  $Z_b(10610)$
and $Z_b(10650)$ in the hidden-bottom decay channels
$\Upsilon(nS)\pi^{\pm}$ ($n=1,2,3$) and $h_b(mP)\pi^\pm$ ($m=1,2$)
of $\Upsilon(5S)$ \cite{Collaboration:2011gj}. The measured
parameters of $Z_b(10610)$ and $Z_b(10650)$ are
\begin{eqnarray*}
M_{Z_b(10610)}/\Gamma_{Z_b(10610)}&=&10608.4\pm2.0/15.6\pm2.5\, \mathrm{MeV},\\
M_{Z_b(10650)}/\Gamma_{Z_b(10650)}&=&10653.2\pm1.5/14.4\pm3.2\,
\mathrm{MeV}.
\end{eqnarray*}
The analysis of the angular distribution indicates that the
quantum numbers of both $Z_b(10610)$ and $Z_b(10650)$ are
$I^{G}(J^P)=1^+(1^+)$. Both $Z_b(10610)$ and $Z_b(10650)$ are
charged hidden-bottom states. Moreover they are very close to the
thresholds of $B\bar{B}^*$ and $B^*\bar{B}^*$
\cite{Nakamura:2010zzi}, respectively. Thus, $Z_b(10610)$ and
$Z_b(10650)$ are ideal candidates of the $B\bar{B}^*$ and
$B^*\bar{B}^*$ S-wave molecular states, which were studied
extensively in Refs. \cite{Liu:2008fh,Liu:2008tn}.

On the other hand, a new puzzle arises in the theoretical study
\cite{Chen:2011qx,Ali:2009es} of the dipion invariant mass
distribution and the $\cos\theta$ distribution of the anomalous
$\Upsilon(2S) \pi^{+} \pi^{-}$ production near the peak of
$\Upsilon(5S)$ \cite{Abe:2007tk}. While all the other calculations
are well in accord with the Belle data, the predicted differential
width $d\Gamma(\Upsilon(5S)\to\Upsilon(2S)\pi^+\pi^-)/d\cos\theta$
disagrees with the Belle measurement \cite{Chen:2011qx}. In this
work, we will illustrate that the inclusion of these two $Z_b$
states in the $\Upsilon(5S) \to \Upsilon(2S) \pi^{+} \pi^{-}$
decays explains the puzzling line shape of
$d\Gamma(\Upsilon(5S)\to\Upsilon(2S)\pi^+\pi^-)/d\cos\theta$ very
naturally.

\begin{figure}[htb]
\centering
\begin{tabular}{ccc}
\scalebox{0.9}{\includegraphics{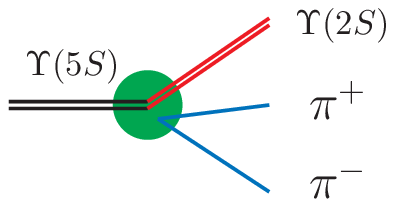}}&\raisebox{4.8ex}{\huge{$+$}}&
\scalebox{0.9}{\includegraphics{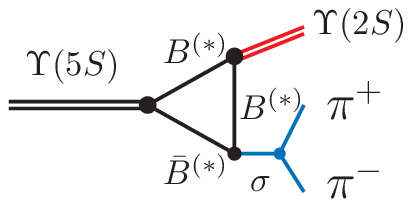}}\\
(a)&&(b)\\
\scalebox{0.8}{\includegraphics{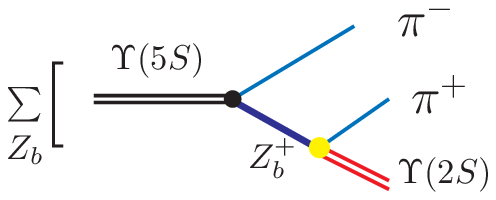}}&\raisebox{4.8ex}{\huge{$+$}}&
\scalebox{0.8}{\includegraphics{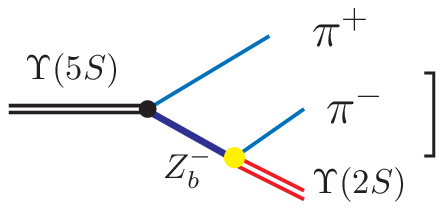}}\\
(c)&&(d)
\end{tabular}
\caption{The diagrams in the $\Upsilon(5S)$ hidden-bottom decay.
Here, Fig. \ref{decay} (a) represents the $\Upsilon(5S)$ direct
decay into $\Upsilon(2S)\pi^+\pi^-$, while Fig. \ref{decay} (b)
denotes the intermediate hadronic loop contribution to
$\Upsilon(5S)\to \Upsilon(2S)\pi^+\pi^-$. (c) and (d) describe the
intermediate $Z_b^{\pm}$ contribution to $\Upsilon(5S)\to
\Upsilon(2S)\pi^+\pi^-$, where
$Z_b^{\pm}=\{Z_b(10610)^\pm,Z_b(10650)^\pm\}$.} \label{decay}
\end{figure}

In general, there exist three mechanisms for the $\Upsilon(5S)$
hidden-bottom decays with the dipion emission
$$\Upsilon(5S)\to
\Upsilon(2S)(p_1)\pi^+(p_2)\pi^-(p_3).$$ The first one is the
$\Upsilon(2S)\pi^+\pi^-$ direct production by $\Upsilon(5S)$ decay
(see Fig. \ref{decay} (a)). The
so-called direct production of
$\Upsilon(5S)\to\Upsilon(2S)\pi^+\pi^-$ denotes that there does
not exist the contribution from the intermediate mesons (such as
$\sigma(600),f_0(980)$, hadronic loop constructed by $B^{(*)}$ or
$B_s^{(*)}$ mesons, $Z_b$) to
$\Upsilon(5S)\to\Upsilon(2S)\pi^+\pi^-$. Thus, the direct
production of $\Upsilon(5S)\to\Upsilon(2S)\pi^+\pi^-$ provides the
background contribution.

The QCD Multipole Expansion (QME) method \cite{Kuang:1981se} is
generally applied to deal with the dipion transition between heavy
quarkonia. So far, there exist many theoretical efforts study the
dipion transitions between the bottomonia
\cite{Yan:1980uh,Kuang:1981se,Zhou:1990ik,Anisovich:1995zu,Guo:2004dt,Guo:2006ai,Simonov:2008qy,Simonov:2008sw}
(see Refs. \cite{Voloshin:1987rp,Besson:1993mm,Kuang:2006me} for a
detailed review). In this work, we do not intend to calculate the
contribution from the direct transition under the framework of the
QME method, but alternatively follow the effective Lagrangian
approach to describe $\Upsilon(5S)\to\Upsilon(2S)\pi^+\pi^-$
transitions. The transition amplitude of the direct production of
$\Upsilon(5S)\to\Upsilon(2S)\pi^+\pi^-$ can be written as
\begin{eqnarray}
&&\mathcal{M}[\Upsilon(5S)\to
\Upsilon(2S)\pi^+\pi^-]_{\mathrm{Direct}}\nonumber\\&&
={\mathcal{F}^{(n)}\over{f_\pi^2}}\epsilon_{\Upsilon(5S)}\cdot
\epsilon_{\Upsilon(2S)}\Big\{\Big[q^2-\kappa^{(n)}(\Delta
M)^2\Big(1+\frac{2m^2_\pi}{q^2}\Big)\Big]_{\mathrm{S-wave}}\nonumber\\&&
\quad+\Big[\frac{3}{2}\kappa^{(n)}\big((\Delta M)^2-q^2\big)
\Big(1-\frac{4m_\pi^2}{q^2}\Big)
\Big(\cos\theta^2-\frac{1}{3}\Big)\Big]_{\mathrm{D-wave}}\Big\}
,\nonumber\\\label{direct}
\end{eqnarray}
which was suggested by Novikov and Shifman in the study of the
$\psi^\prime\to J/\psi\pi^+\pi^-$ decay \cite{Novikov:1980fa},
where the subscripts S-wave and D-wave denote the S-wave and
D-wave contributions respectively. $\Delta M$ is the mass
difference between $\Upsilon(5S)$ and $\Upsilon(2S)$.
$q^2=(p_2+p_3)^2\equiv m_{\pi^+\pi^-}^2$ denotes the invariant
mass of $\pi^+\pi^-$, while $\theta$ is the angle between
$\Upsilon(5S)$ and $\pi^-$ in the $\pi^+\pi^-$ rest frame. The
pion decay constant and mass are taken as $f_\pi=130$ MeV and
$m_\pi=140$ MeV, respectively. In Eq. (\ref{direct}), $\kappa$ and
$\mathcal{F}$ are free parameters to be determined when fitting
the experimental data.

Different from the other low-lying bottomonia with
$J^{PC}=1^{--}$, $\Upsilon(5S)$ is above the
$B^{(*)}\bar{B}^{(*)}$ thresholds and predominantly decays into
$B^{(*)}\bar{B}^{(*)}$ pair, which may render the coupled channel
effect quite important
\cite{Meng:2007tk,Meng:2008dd,Simonov:2008ci}. When exploring the
$\Upsilon(5S)$ hidden-bottom decay, the coupled channel effect has
to be taken into account. In other words, there also exists the
second mechanism contributing to the $\Upsilon(5S)\to
\Upsilon(2S)\pi^+\pi^-$ transitions as shown in Fig. \ref{decay}
(b), where the intermediate $B^{(*)}$ and $\bar{B}^{(*)}$ hadronic
loop is the bridge to connect the initial state $\Upsilon(5S)$ and
final state $\Upsilon(2S)\pi^+\pi^-$. Furthermore,
$\Upsilon(5S)\to \Upsilon(nS)\pi^+\pi^-$ can be approximately
expressed as a sequential decay process. $\Upsilon(5S)$ first
transits into $\Upsilon(2S)$ and the scalar meson $\sigma(600)$.
Then $\sigma(600)$ couples with the dipion. Choosing $\sigma(600)$
as the intermediate state contribution to the $\Upsilon(5S)\to
\Upsilon(2S)\pi^+\pi^-$ process is not only consistent with the
Belle data \cite{Collaboration:2011gj,Abe:2007tk} but also allowed by the phase space
of the decay channel.

If comparing the dipion invariant mass spectrum of
$\Upsilon(5S)\to\Upsilon(2S)\pi^+\pi^-$ in Refs.
\cite{Collaboration:2011gj,Abe:2007tk}, the data in Ref.
\cite{Collaboration:2011gj} at the higher end of $m_{\pi^+\pi^-}$
are qualitatively different from those in Ref. \cite{Abe:2007tk},
where the total events in Ref. \cite{Abe:2007tk} are at least one
order of magnitude less than those in Ref.
\cite{Collaboration:2011gj}. Such a large accumulation of events
at $m_{\pi^+\pi^-}>700$ MeV \cite{Collaboration:2011gj} might be
due to the contribution from the tail of the intermediate
$f_0(980)$. We did not include the $f_0(980)$ contribution when we
analyzed the data in \cite{Abe:2007tk}. Considering the situation
of the new data of the dipion invariant mass spectrum of
$\Upsilon(5S)\to\Upsilon(2S)\pi^+\pi^-$
\cite{Collaboration:2011gj}, we also include $f_0(980)$
contribution to the analysis of
$\Upsilon(5S)\to\Upsilon(2S)\pi^+\pi^-$ in the following.

The effective Lagrangians relevant to the Fig. \ref{decay} (b)
include
\begin{eqnarray}
{\cal L}_{\Upsilon\cal{BB}}&=&i g_{\Upsilon \cal{BB}}\Upsilon_{\mu}(\partial^{\mu}\cal{BB}^{\dag}-\cal{B\partial^{\mu}B}^{\dag}),\\
{\cal L}_{\Upsilon\cal{B^*B}}&=&-i g_{\Upsilon \cal{B^*B}}\varepsilon^{\mu\nu\alpha\beta}\partial_{\mu}\Upsilon_{\nu}(\partial_{\alpha}\cal{B}_{\beta}^*\cal{B}^{\dag}+\cal{B\partial_{\alpha}B}^{*\dag}_{\beta}),\\
{\cal L}_{\Upsilon\cal{B^*B^*}}&=&-i g_{\Upsilon {\cal
B^*B^*}}\{\Upsilon^{\mu}(\partial_{\mu}{\cal
B^{*\nu}B^{*\dag}_{\nu}}-{\cal B^{*\nu}}\partial_{\mu}{\cal
B^{*\dag}_{\nu}})\nonumber\\&&+
(\partial_{\mu}\gamma_{\nu}{\cal B^{*\nu}}-\Upsilon_{\nu}\partial_{\mu}{\cal B^{*\nu}})\cal{B^{*\mu\dag}}\nonumber\\
&&+{\cal B^{*\mu}}(\Upsilon^{\nu}\partial_{\mu}{\cal B}^{*\dag}_
{\nu}-\partial_{\mu}\Upsilon_{\nu}\cal{B}^{*\nu\dag})\},
\end{eqnarray}
and
\begin{eqnarray}
{\cal L}_{\cal SB^{(*)}B^{(*)}}=g_{\cal BBS}{\cal
SBB^{\dag}}-g_{\cal B^*B^*S}{\cal SB^*B^{*\dag}}
\end{eqnarray}
where  ${\cal B}=(\bar{B^0},B^-,B_s^-)$ and $({\cal
B}^{\dag})^{T}=(B^0,B^+,B_s^+)$. There are 4 diagrams. Thus, the
concrete expressions of decay amplitudes are written as
\begin{eqnarray}
\nonumber\mathcal{M}_{B\bar
B}^{B}&=&(i)^3\int\frac{d^4q}{(2\pi)^4}[ig_{\Upsilon(5S)BB}\epsilon^\mu_{\Upsilon(5S)}
(ip_{2\mu}-ip_{1\mu})]\nonumber\\&&\times[ig_{\Upsilon(nS)BB}
\epsilon^{\rho}_{\Upsilon(nS)}(-ip_{1\rho}-iq_{\rho})][g_{BBS}]\nonumber\\
&&\times\frac{1}{p_1^2-m_B^2}\frac{1}{p_2^2-m_B^2}\frac{1}{q^2-m_B^2}\mathcal{F}(q^2),\\
\nonumber\mathcal{M}_{B\bar
B^\ast}^{B^\ast}&=&(i)^3\int\frac{d^4q}{(2\pi)^4}[-g_{\Upsilon(5S)BB^\ast}\varepsilon_{\mu\nu\alpha\beta}(-ip^{\mu}_{0})\epsilon^{\nu}_{\Upsilon(5S)}(ip^\alpha_{2})]
\nonumber\\&&\times[-g_{\Upsilon(nS)BB}\varepsilon_{\delta\tau\theta\phi}(ip^\delta_{3})\epsilon^\tau_{\Upsilon(nS)}(iq^{\theta})][-g_{B^*B^*S}]\nonumber
\\&&\times\frac{1}{p_1^2-m_B^2}\frac{-g^{\beta\rho}+p^\beta_2p^\rho_2/m_{B^*}^{2}}{p^2_2-m_{B^*}^{2}}\frac{-g^{\phi\rho}+q^\phi q^\rho/m_{B^*}^{2}}{q^2_2-m_{B^*}^{2}}
\mathcal{F}(q^2),\nonumber\\\\
\nonumber\mathcal{M}_{B\bar
B^\ast}^{B}&=&(i)^3\int\frac{d^4q}{(2\pi)^4}[-g_{\Upsilon(5S)B^*B}\varepsilon_{\mu\nu\alpha\beta}(-ip^{\mu}_{0})\epsilon^{\nu}_{\Upsilon(5S)}(ip^\alpha_{1})]
\nonumber\\&&\times[-g_{\Upsilon(nS)B^{*}B}\varepsilon_{\delta\tau\theta\phi}(ip^\delta_{3})\epsilon^\tau_{\Upsilon(nS)}(-ip^{\theta}_{1})][g_{BBS}]
\nonumber\\&&\times\frac{-g^{\beta\phi}+p^\beta_1p^\phi_1/m_{B^*}^{2}}{p^2_1-m_{B^*}^{2}}\frac{1}{p^2_2-m_{B}^{2}}
\frac{1}{q^2-m_B^2}\mathcal{F}(q^2),\\
\nonumber\mathcal{M}_{B^\ast\bar
B^\ast}^{B^\ast}&=&(i)^3\int\frac{d^4q}{(2\pi)^4}[-ig_{\Upsilon(5S)B^*B^*}\epsilon^{\mu}_{\Upsilon(5S)}((ip_{2\mu}-ip_{1\mu}
)g_{\nu\rho}\nonumber\\&&+(-ip_{0\rho}-ip_{2\rho})g_{\mu\nu}+(ip_{1\nu}+ip_{0\nu})g_{\mu\rho})]\nonumber\\
&&\times[-ig_{\Upsilon(nS)B^*B^*}\epsilon^{\phi}_{\Upsilon(nS)}((-ip_{1\phi}-iq_{\phi}
)g_{\alpha\beta}\nonumber\\&&+(ip_{3\beta}+ip_{1\beta})g_{\alpha\phi}+(iq_{\alpha}-ip_{3\alpha})g_{\beta\phi})]
[-g_{B^*B^*S}]\nonumber\\
&&\times\frac{-g^{\rho\alpha}+p^\rho_1p^\alpha_1/m_{B^*}^{2}}{p^2_1-m_{B^*}^{2}}
\frac{-g^{\nu\tau}+p^\nu_2p^\tau_2/m_{B^*}^{2}}{p^2_2-m_{B^*}^{2}}\nonumber\\&&\times
\frac{-g^{\beta\tau}+q^\beta q^\tau/m_{B^*}^{2}}{q^2-m_{B^*}^{2}}
\mathcal{F}(q^2).
\end{eqnarray}
The amplitude $\mathcal{M}_{AB}^{C}$ indicates that the initial
$\Upsilon(5S)$ dissolves into intermediate $AB$, which transit
into the final $\Upsilon(2S)$ and scalar meson by exchanging meson
$C$. In the above expressions, the form factor is introduced by
$\mathcal{F}(q^2)=(\Lambda^2-m_E^2)/(q^2-m_E^2)$. And $m_E$ is the
mass of the exchanged $B^{(*)}$ meson in the
$B^{(*)}\bar{B}^{(*)}\to \Upsilon(2S)\mathcal{S}$ transitions
shown in Fig. \ref{decay} (b) and $\Lambda=m_E+ \alpha
\Lambda_{QCD}$ with $\Lambda_{QCD}=220$ MeV. As indicated in Ref.
\cite{Chen:2011qx}, we can parameterize the decay amplitude of
$\Upsilon(5S)\to \Upsilon(2S)\pi^+\pi^-$ corresponding to Fig.
\ref{decay} (b) as
\begin{eqnarray}
&&\mathcal{M}[\Upsilon(5S)\to \Upsilon(2S)\sigma(600)\to
\Upsilon(2S)\pi^+\pi^-]\nonumber\\
&&=
\frac{\epsilon_{\Upsilon(5S)}\cdot \epsilon_{\Upsilon(2S)}^*F_\sigma}{(p_2+p_3)^2-m_{\sigma}^2 + i m_{\sigma}
\Gamma_{\sigma}}, \label{ha1}
\end{eqnarray}
if only considering the S-wave contribution. Here, we introduce
$F_\sigma$ as the fitting parameter.

Similar to Eq. (\ref{ha1}), the parameterized decay amplitude of
$\Upsilon(5S)\to\Upsilon(2S)\pi^+\pi^-$ with $f_0(980)$ as the
intermediate state can be expressed as
\begin{eqnarray}
&&\mathcal{M}[\Upsilon(5S)\to \Upsilon(2S)f_0(980)\to
\Upsilon(2S)\pi^+\pi^-]\nonumber\\
&&=  \frac{\epsilon_{\Upsilon(5S)}\cdot
\epsilon_{\Upsilon(2S)}^*F_{f_0}}{(p_2+p_3)^2-m_{f_0}^2 + i
m_{f_0} \Gamma_{f_0}}, \label{haf0}
\end{eqnarray}
which corresponds to Fig. \ref{decay} (b) with the replacement
$\sigma\to f_0(980)$.

Regarding the contribution of these two newly observed $Z_b$
states to the $\Upsilon(5S)\to \Upsilon(2S)\pi^+\pi^-$ process, we
introduce the third mechanism depicted in Fig. \ref{decay} (c) and
(d), where $Z_b^\pm$s are the intermediate states and interact
with $\Upsilon(5S)\pi^{\mp}$ and $\Upsilon(2S)\pi^{\pm}$. The
general expressions of the amplitudes of Fig. \ref{decay} (c) and
(d) are
\begin{eqnarray}
&&\mathcal{M}[\Upsilon(5S)\to Z_b^+\pi^-\to
\Upsilon(2S)\pi^+\pi^-]_{Z_b^+}\nonumber\\
&&= F_{Z_b^+} \epsilon_{\Upsilon(5S)}^\mu
\epsilon_{\Upsilon(2S)}^{*\nu} \frac{-g_{\mu \nu} + (p_1^\mu
+p_2^\mu) (p_1^\nu +p_2^\nu)/m_{Z_b}^2}{ (p_1+p_2)^2-m_{Z_b}^2 +
im_{Z_b}
\Gamma_{Z_b}}\\
&&\mathcal{M}[\Upsilon(5S)\to Z_b^-\pi^+\to
\Upsilon(2S)\pi^-\pi^+]_{Z_b^-}\nonumber\\
&&= F_{Z_b^-}\epsilon_{\Upsilon(5S)}^\mu
\epsilon_{\Upsilon(2S)}^{*\nu} \frac{-g_{\mu \nu} + (p_1^\mu
+p_3^\mu) (p_1^\nu +p_3^\nu)/m_{Z_b}^2}{ (p_1+p_3)^2-m_{Z_b}^2 +
im_{Z_b} \Gamma_{Z_b}},
\end{eqnarray}
respectively, where we define $F_{Z_b^+}=g_{_{\Upsilon(5S)Z_b^+
\pi}}g_{_{Z_b^+ \Upsilon(2S) \pi^+}}$ and
$F_{Z_b^-}=g_{_{\Upsilon(5S)Z_b^- \pi}}g_{_{Z_b^- \Upsilon(2S)
\pi^-}}$. Since Fig. \ref{decay} (c) and (d) are related to each
other by charge-conjugation, thus $F_{Z_b^-}=F_{Z_b^+}=F_{Z_b}$.

Thus, the total decay amplitude of the
$\Upsilon(5S)\to \Upsilon(2S)\pi^+\pi^-$ decay is
\begin{eqnarray}
\mathcal{M}_{\mathrm{total}}&=&\mathcal{M}[\Upsilon(5S)\to
\Upsilon(2S)\pi^+\pi^-]_{\mathrm{Direct}}\nonumber\\&&+e^{i\phi_{\sigma}}
\mathcal{M}[\Upsilon(5S)\to \Upsilon(2S)\sigma(600)\to
\Upsilon(2S)\pi^+\pi^-]\nonumber\\&& \nonumber\\&&+e^{i\phi_{f_0}}
\mathcal{M}[\Upsilon(5S)\to \Upsilon(2S)f_0(980)\to
\Upsilon(2S)\pi^+\pi^-]\nonumber\\&&+\sum_{_{Z_b}}e^{i\varphi_{_{Z_b}}}\bigg\{
\mathcal{M}[\Upsilon(5S)\to Z_b^+\pi^-\to
\Upsilon(2S)\pi^+\pi^-]_{Z_b^+}\nonumber\\&&+\mathcal{M}[\Upsilon(5S)\to
Z_b^-\pi^+\to \Upsilon(2S)\pi^+\pi^-]_{Z_b^-}\bigg\},\label{total}
\end{eqnarray}
where we have introduced the phase angles $\phi_\sigma$, $\phi_{f_0}$,
$\varphi_{_{Z_b(10610)}}$ and $\varphi_{_{Z_b(10650)}}$.

As a three body decay, the differential decay width for
$\Upsilon(5S) \to \Upsilon(2S) \pi^+ \pi^-$ read as,
\begin{eqnarray}
d\Gamma =\frac{1}{3} \frac{1}{(2 \pi)^3} \frac{1}{32
m_{\Upsilon(5S)}^3} {|\mathcal{M}_{\mathrm{total}}|^2}
dm_{\Upsilon(2S) \pi}^2 dm_{\pi\pi}^2
\end{eqnarray}
with $m_{\Upsilon(2S) \pi^+}^2 = (p_1 + p_2)^2$ and
$m_{\pi^+\pi^-}^2 =(p_2 +p_3)^2$. The relevant resonance
parameters are listed in Table. \ref{Tab-Input}.
\begin{table}[htb]
\centering %
\caption{The resonance parameters adopted in our calculation
\cite{Nakamura:2010zzi,Aitala:2000xu,Collaboration:2011gj}.
\label{Tab-Input}}
\begin{tabular}{cccccc}
 \toprule[1pt]
State&Mass (GeV)&State&Mass (GeV)&Width (GeV)\\\midrule[1pt]
$\Upsilon(5S)$&10.870&$\sigma(600)$  & 0.478  & 0.324\\
&&$f_0(980)$& 0.980&0.100\\
$\Upsilon(2S)$&10.023&$Z_{b}(10610)$ & 10.608 & 0.0156\\
              &      &$Z_{b}(10650)$ & 10.653 & 0.0144\\
 \bottomrule[1pt]
\end{tabular}
\end{table}

\begin{figure}[htb]
\centering \scalebox{0.7}{\includegraphics{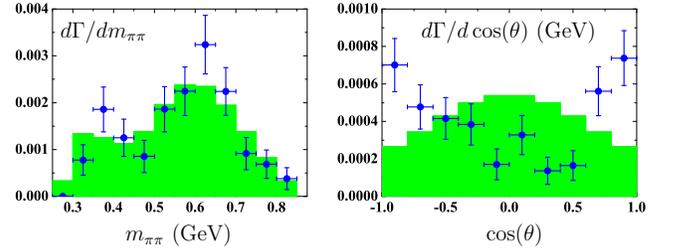}}
\caption{(Color online). The dipion invariant mass ($m_{\pi\pi}$)
distribution (left-panel) and the $\cos \theta$ distribution
(right-panel) of the $\Upsilon(5S)\to \Upsilon(2S) \pi^{+}\pi^{-}$
decay. The dots with error bars are the results measured by Belle
\cite{Abe:2007tk}, while the green histograms is the best fit from
our model without including the intermediate $Z_b(10610)^\pm$ and
$Z_b(10650)^\pm$ contribution to $\Upsilon(5S)\to \Upsilon(2S)
\pi^{+}\pi^{-}$. When fitting the experimental data \cite{Abe:2007tk}, we only
include $\sigma(600)$ contribution.
 \label{fit1}}
\end{figure}

If considering only the contributions from Fig. \ref{decay} (a)
and (b) in our present scenario, we have four free parameters as
listed in Table \ref{withouta}, where the $\sigma(600)$ contribution is included to fit the Belle data \cite{Abe:2007tk}. With the help of the MINUIT
package, we perform the global fit to the experimental data of the
dipion invariant mass spectrum distribution and the $\cos\theta$
distribution of the $\Upsilon(2S)\pi^+\pi^-$ production near the peak
of $\Upsilon(5S)$ \cite{Abe:2007tk}. The best fit to the dipion
invariant mass spectrum distribution is shown in the left-panel in
Fig. \ref{fit1}. Unfortunately the corresponding $\cos\theta$
distribution of the $\Upsilon(2S)\pi^+\pi^-$ production strongly
deviates from the Belle data as shown in the right-panel of Fig.
\ref{fit1}. The values of the obtained fitting parameters are
presented in Table \ref{withouta}. Such discrepancy between
theoretical and experimental results stimulates a {\it New Puzzle}
first indicated in Ref. \cite{Chen:2011qx}. At present, solving
these new puzzle becomes an important and intriguing research
topic, which will be helpful to underlying mechanism behind the
$\Upsilon(5S)\to \Upsilon(2S)\pi^+\pi^-$ decay.

\begin{table}[htb]
\centering %
\caption{The values of the fitting parameters for the best fit to
the Belle data of $\Upsilon(2S) \pi^+ \pi^-$ production near the
peak of $\Upsilon(5S)$ \cite{Abe:2007tk} without considering the
contributions from $Z_b(10610)$ and $Z_b(10650)$. For the obtained
central values of the parameters, the corresponding partial decay
width of $\Upsilon(5S) \to \Upsilon(2S) \pi^+ \pi^-$ is $0.836$
MeV. \label{withouta}}
\begin{tabular}{cccc}
 \toprule[1pt]
Parameter &  Value & Parameter &  Value \\
 \midrule[1pt]
$\mathcal{F}$  &  $ 0.943 \pm 0.071$  & $\kappa$ & $0.739 \pm 0.034$\\
$F_{\sigma}$   &  $25.603 \pm 2.175\ \mathrm{GeV}^2$ &
$\phi_{\sigma}$  & $  2.623 \pm 0.132 $ Rad\\
 \bottomrule[1pt]
\end{tabular}
\end{table}

\begin{figure}[htb]
\centering \scalebox{0.7}{\includegraphics{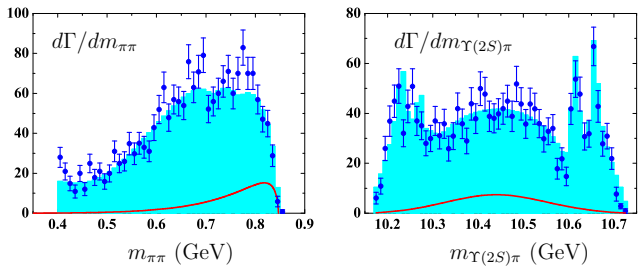}}
\caption{(Color online). The invariant mass spectra of $\pi^+\pi^-$
and $\Upsilon(2S)\pi^+$ for $\Upsilon(5S)\to
\Upsilon(2S)\pi^+\pi^-$. Here, the histograms are theoretical
results obtained in our scenario including the intermediate
$f_0(980)$ contribution, while dots with error bars are the Belle data
in Ref. \cite{Collaboration:2011gj}. We also plot the $f_0(980)$
contribution separately (red solid lines).
 \label{fitf0}}
\end{figure}

In contrast, we consider the contribution from $Z_b(10610)$ and
$Z_b(10650)$ in the following and discuss the dependence of
$d\Gamma/dm_{\pi^+\pi^-}$ and $d\Gamma/d\cos\theta$ of
$\Upsilon(5S)\to \Upsilon(2S)\pi^+\pi^-$ on $m_{\pi^+\pi^-}$ and
$\cos\theta$ respectively. Under this scheme, we refit the Belle
data \cite{Collaboration:2011gj} with Eq. (\ref{total}). There are 10
fitting parameters as listed in Table \ref{with}. In Fig.
\ref{fitf0}, we present a comparison between the Belle data (dots with
error bars) and our best fit (histograms) to the Belle data \cite{Collaboration:2011gj}, which
indicates that the line shapes of the invariant mass spectra of $\pi^+\pi^-$ and
$\Upsilon(2S)\pi^+$ for $\Upsilon(5S)\to \Upsilon(2S)\pi^+\pi^-$
describe the Belle data \cite{Collaboration:2011gj} well. The
double-peak structure around $10.6$ GeV and its reflection around
10.25 GeV are reproduced by our model well. With the
central values of these parameters in Table
\ref{with}, we obtain the partial decay
width of $\Upsilon(5S) \to \Upsilon(2S) \pi^+ \pi^-$
$\Gamma=0.915$ MeV, which is consistent with the Belle measurement
$\Gamma=0.85\pm0.07(\mathrm{stat})\pm0.16(\mathrm{syst})$ MeV
\cite{Abe:2007tk}. Thus, the contribution from these charged $Z_b$
resonances provides a possible solution to the puzzle why the
$\Upsilon(5S)\to \Upsilon(2S)\pi^+\pi^-$ decay width is abnormally
large \cite{Abe:2007tk}.

\begin{table}[htb]
\centering %
\caption{The values of the fitting parameters for the
$\Upsilon(5S) \to \Upsilon(2S) \pi^+ \pi^-$ decay after including
the contributions from $Z_b(10610)$ and $Z_b(10650)$.
\label{with}}
\begin{tabular}{cccc}
 \toprule[1pt]
Parameter &  Value & Parameter &  Value \\
 \midrule[1pt]
%
%
$\mathcal{F}$  &  $ 1.404 \pm 0.068$  &
$\kappa$ & $0.301 \pm 0.013$\\
%
%
$F_{\sigma}$   &  $20.037 \pm 0.423\ \mathrm{GeV}^2$ &
$\phi_{\sigma}$  & $0.907 \pm 0.132 $ rad\\
%
%
$F_{f_0}$   &  $17.076 \pm 3.563 \ \mathrm{GeV}^2$ &
$\phi_{f_0}$  & $-0.753 \pm 0.140 $ rad\\
%
%
$F_{Z_b(10610)}$   &  $3.412 \pm 0.385\ \mathrm{GeV}^2$ &
$\varphi_{_{Z_b(10610)}}$  & $  -3.135 \pm 0.030 $ rad\\
%
%
$F_{Z_b(10650)}$   &  $2.994 \pm 0.261 \ \mathrm{GeV}^2$ &
$\varphi_{_{Z_b(10650)}}$  & $  -2.836 \pm 0.165 $ rad\\
 \bottomrule[1pt]
\end{tabular}
\end{table}

\begin{figure}[htb]
\centering \scalebox{0.7}{\includegraphics{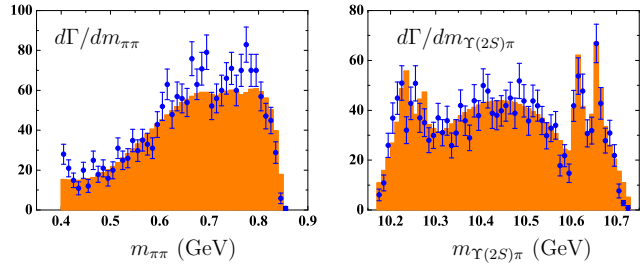}}
\caption{(Color online). The
distribution invariant mass spectra $m_{\pi \pi}$ and
$m_{\Upsilon(2S) \pi}$ for $\Upsilon(5S) \to \Upsilon(2S) \pi^+
\pi^-$ without including the contributions from $f_0(980)$. Here,
we use Eq. (\ref{total}) to redo the analysis. The histograms are
the fitting results. The dots with errors correspond to the Belle
data \cite{Collaboration:2011gj}. \label{h1} }
\end{figure}

From Table \ref{with}, we notice that the
uncertainty of $F_{f_0}$ is one order-of-magnitude larger than
that of $F_\sigma$, which means the fit is less sensitive to the
$f_0(980)$ than to the $\sigma(600)$.
Using Eq. (\ref{total}), we reanalyze the new Belle data in Ref.
\cite{Collaboration:2011gj} with the obtained fitting parameters
in Table \ref{without}, where we do not include the $f_0(980)$
contribution. The comparison between our fitting result and the
experimental data are given in Fig. \ref{h1}. By the scenario in
Eq. (\ref{total}), we reproduce the Belle data well, which
confirms that the intermediate $f_0(980)$ contribution to
$\Upsilon(5S)\to \Upsilon(2S)\pi^+\pi^-$ is small. If comparing
the obtained values of the fitting parameter in Tables \ref{with}
and \ref{without}, we notice that the eight common parameters do
not change much in the two schemes. With the parameters listed in
Tables \ref{with} and \ref{without}, we also present the
$\cos\theta$ distribution with and without the intermediate $f_0$
contribution. The experimental measurement of the $\cos\theta$
distribution for $\Upsilon(5S)\to \Upsilon(2S)\pi^+\pi^-$ \cite{Abe:2007tk} can be
described well with the scenarios in this work. This fact
indicates that the two $Z_b$ structures play important role in the
understanding of the Belle data, especially the $\cos\theta$
distribution of $\Upsilon(5S)\to \Upsilon(2S)\pi^+\pi^-$.

\begin{table}[htb]
\centering %
\caption{The values of the fitting
parameters for the $\Upsilon(5S) \to \Upsilon(2S) \pi^+ \pi^-$
decay after including the contributions from $Z_b(10610)$ and
$Z_b(10650)$. These parameters are obtained by fitting the new
experimental data without including the contributions from
$f_0(980)$.\label{without}}
\begin{tabular}{cccc}
 \toprule[1pt]
Parameter &  Value & Parameter &  Value \\
 \midrule[1pt]
%
%
$\mathcal{F}$  &  $ 1.073 \pm 0.064$  &
$\kappa$ & $0.379 \pm 0.034$\\
%
%
$F_{\sigma}$   &  $23.833 \pm 2.503\ \mathrm{GeV}^2$ &
$\phi_{\sigma}$  & $1.127 \pm 0.128 $ rad\\
%
%
$F_{Z_b(10610)}$   &  $3.200 \pm 0.345\ \mathrm{GeV}^2$ &
$\varphi_{_{Z_b(10610)}}$  & $  -3.141 \pm 0.076 $ rad\\
%
%
$F_{Z_b(10650)}$   &  $2.686 \pm 0.306 \ \mathrm{GeV}^2$ &
$\varphi_{_{Z_b(10650)}}$  & $  -2.703 \pm 0.225 $ rad\\
 \bottomrule[1pt]
\end{tabular}
\end{table}

\begin{figure}[htb]
\centering \scalebox{0.7}{\includegraphics{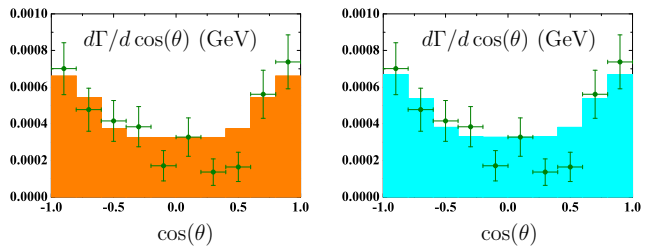}}
\caption{(Color online). The
$\cos\theta$ distributions for $\Upsilon(5S) \to \Upsilon(2S)
\pi^+ \pi^-$. The histograms in the left-hand side and right-hand
side diagrams are the fitting results
 without and with the contribution from $f_0(980)$. The dots with errors correspond to the Belle data \cite{Abe:2007tk}. \label{ha} }
\end{figure}

In summary, the Belle Collaboration announced an exciting
observation of two charged bottomonium-like states $Z_b(10610)$
and $Z_b(10650)$. These $Z_b$ states are good candidates of exotic
states, which calls for theoretical efforts in revealing their
underlying structures. Carrying out the phenomenological study
relevant to $Z_b(10610)$ and $Z_b(10650)$ is one of the important
and valuable issues of heavy quarkonium physics, which is full of
challenges and opportunities
\cite{Brambilla:2004wf,Brambilla:2010cs}.

The $Z_b(10610)$ and $Z_b(10650)$ states are related to the
anomalous phenomena of $\Upsilon(2S)\pi^+\pi^-$ production near
$\Upsilon(5S)$ previously reported by Belle \cite{Abe:2007tk}.
Comparing the fitting results without and with the contributions
from the newly observed states, we notice that the intermediate
$Z_b(10610)$ and $Z_b(10650)$ play a crucial role in the behavior
of $d\Gamma(\Upsilon(5S) \to \Upsilon(2S) \pi^+
\pi^-)/d\cos\theta$. The inclusion of the $Z_b(10610)$ and
$Z_b(10650)$ contribution to $\Upsilon(5S)\to
\Upsilon(2S)\pi^+\pi^-$ provides a unique mechanism of understand
the puzzling $\cos\theta$ distribution of $\Upsilon(2S)\pi^+\pi^-$
production near $\Upsilon(5S)$ \cite{Abe:2007tk}. The double-peak
structure and its reflection in the $\Upsilon(2S)\pi^+$ invariant
mass spectrum of $\Upsilon(5S)\to \Upsilon(2S)\pi^+\pi^-$
\cite{Collaboration:2011gj} are also reproduced by this mechanism.
In this work, the values of the
fitting parameters in our scenario are obtained by fitting Belle
data \cite{Collaboration:2011gj,Abe:2007tk}. To some extent, the
interpretation of the values of these parameters is related to the
understanding of background, the structures of two $Z_b$ states
etc, which is an interesting research topic.

Besides finding the signals of $Z_b(10610)$ and $Z_b(10650)$ in
$\Upsilon(2S)\pi^\pm$ decay channel, Belle's analysis of its
remaining four hidden-bottom decay channels
$\Upsilon(nS)\pi^{\pm}$ ($n=1,3$) and $h_b(mP)\pi^\pm$ ($m=1,2$)
also indicate the observation of $Z_b(10610)$ and $Z_b(10650)$
\cite{Collaboration:2011gj}. The present formalism can be extended
to study the dipion invariant mass distribution and the
$\cos\theta$ distribution of $\Upsilon(5S)\to
\Upsilon(1S,3S)\pi^{+}\pi^{-}$ and $\Upsilon(5S)\to
h_b(1P,2P)\pi^{+}\pi^{-}$ decay.

Additionally, Belle's measurement favors the $B\bar{B}^*$ and
$B^*\bar{B}^*$ molecular explanation of the $Z_b(10610)$ and
$Z_b(10650)$ resonances respectively. The possible S-wave
$B\bar{B}^*$ and $B^*\bar{B}^*$ molecular states were investigated
extensively in Refs. \cite{Liu:2008fh,Liu:2008tn}. Very recently,
the authors in Ref. \cite{Bondar:2011ev} discussed the special
decay behaviour of the J=1 S-wave $B\bar{B}^*$ and $B^*\bar{B}^*$
molecular states based on the heavy quark symmetry. Future
dynamical study of the mass and decay pattern of the S-wave
$B\bar{B}^*$ and $B^*\bar{B}^*$ molecular states are very
desirable.

\vfil
\section*{Acknowledgment}

This project is supported by the National Natural Science
Foundation of China under Grants No. 11175073, No. 11005129,
No. 11035006, No. 11047606, No. 11075004, No. 11021092; the Ministry of Education
of China (FANEDD under Grant No. 200924, DPFIHE under Grant No.
20090211120029, NCET under Grant No. NCET-10-0442, the Fundamental
Research Funds for the Central Universities); and the West
Doctoral Project of the Chinese Academy of Sciences.

\end{document}